\documentclass{emulateapj}

\usepackage{lscape}
\usepackage{apjfonts}
\usepackage{amsmath}
\usepackage{graphicx}
\newcommand{\etal}{et al.}

\begin{document}

%\title{The Small Scale Clustering of Quasars at Redshift $z\gtrsim 3$}
\title{Binary Quasars at High Redshift II: Sub-Mpc Clustering at $z\sim 3-4$ }

%BINARY QUASARS AT HIGH REDSHIFT II: SUB-MPC CLUSTERING AT $Z\SIM 3-4$

\shorttitle{SMALL-SCALE QUASAR CLUSTERING AT $z\sim 3-4$}

\shortauthors{SHEN ET AL.}
\author{Yue Shen\altaffilmark{1}, Joseph F. Hennawi\altaffilmark{2,3,4,5},
Francesco Shankar\altaffilmark{6}, Adam D.
Myers\altaffilmark{5,7}, Michael A. Strauss\altaffilmark{1}, S. G.
Djorgovski\altaffilmark{8}, Xiaohui Fan\altaffilmark{9}, Carlo
Giocoli\altaffilmark{10}, Ashish Mahabal\altaffilmark{8}, Donald
P. Schneider\altaffilmark{11}, David H. Weinberg\altaffilmark{12}}

%\author{Yue Shen\altaffilmark{1}, Michael A. Strauss\altaffilmark{1},
%Nicholas P. Ross\altaffilmark{2}, Patrick B. Hall\altaffilmark{3},
%Yen-Ting Lin\altaffilmark{1}, Gordon T. Richards\altaffilmark{4},
%Donald P. Schneider\altaffilmark{2}, David H.
%Weinberg\altaffilmark{5}, Andrew J. Connolly\altaffilmark{6},
%Xiaohui Fan\altaffilmark{7}, Joseph F. Hennawi\altaffilmark{8},
%Francesco Shankar\altaffilmark{5}, Daniel E. Vanden
%Berk\altaffilmark{2}, Neta A. Bahcall\altaffilmark{1}, Robert J.
%Brunner\altaffilmark{9}}

\altaffiltext{1}{Princeton University Observatory, Princeton, NJ
08544.}

\altaffiltext{2}{Department of Astronomy, Campbell Hall,
University of California, Berkeley, California 94720.}

\altaffiltext{3}{NSF Astronomy and Astrophysics Postdoctoral
Fellow.}

\altaffiltext{4}{Max-Planck-Institut f\"{u}r Astronomie,
K\"{o}nigstuhl 17, D-69117 Heidelberg, Germany.}

\altaffiltext{5}{Visiting Astronomer, Kitt Peak National
Observatory, National Optical Astronomy Observatory, which is
operated by the Association of Universities for Research in
Astronomy (AURA) under cooperative agreement with the National
Science Foundation.}

\altaffiltext{6}{Max-Planck-Instit\"{u}t f\"{u}r Astrophysik,
Karl-Schwarzschild-Str. 1, D-85748, Garching, Germany.}

\altaffiltext{7}{Department of Astronomy, University of Illinois
at Urbana-Champaign, Urbana, IL 61801.}

\altaffiltext{8}{Division of Physics, Mathematics, and Astronomy,
California Institute of Technology, Pasadena, CA 91125.}

\altaffiltext{9}{Steward Observatory, 933 North Cherry Avenue,
Tucson, AZ 85721.}

\altaffiltext{10}{Institut\"{u}t f\"{u}r Theoretische Astrophysik,
Zentrum f\"{u}r Astronomie der Universit\"{a}t Heidelberg
Albert-Ueberle-Str. 2, 69120 Heidelberg, Germany.}

\altaffiltext{11}{Department of Astronomy and Astrophysics, 525
Davey Laboratory, Pennsylvania State University, University Park,
PA 16802.}

\altaffiltext{12}{Astronomy Department, Ohio State University,
Columbus, OH 43210.}

%\altaffiltext{2}{Department of Astronomy and Astrophysics, 525
%Davey Laboratory, Pennsylvania State University, University Park,
%PA 16802.}
%
%\altaffiltext{3}{Dept. of Physics \& Astronomy, York University,
%4700 Keele St., Toronto, ON, M3J 1P3, Canada.}
%
%\altaffiltext{4}{Department of Physics, Drexel University, 3141
%Chestnut Street, Philadelphia, PA 19104.}
%
%\altaffiltext{5}{Astronomy Department, Ohio State University,
%Columbus, OH 43210.}
%
%\altaffiltext{6}{Department of Astronomy, University of
%Washington, Box 351580, Seattle, WA 98195.}
%
%\altaffiltext{7}{Steward Observatory, 933 North Cherry Avenue,
%Tucson, AZ 85721.}
%
%\altaffiltext{8}{Department of Astronomy, Campbell Hall,
%University of California, Berkeley, California 94720.}
%
%\altaffiltext{9}{Department of Astronomy, MC-221, University of
%Illinois, 1002 West Green Street, Urbana, IL 61801.}

\begin{abstract}
We present measurements of the small-scale ($0.1\lesssim r\lesssim
1\ h^{-1}$Mpc) quasar two-point correlation function at $z>2.9$,
for a flux-limited ($i<21$) sample of 15 binary quasars compiled
by Hennawi \etal\ (2009). The amplitude of the small-scale
clustering increases from $z\sim 3$ to $z\sim 4$. The small-scale
clustering amplitude is comparable to or lower than power-law
extrapolations (with slope $\gamma=2$) from the large-scale
correlation function of the $i<20.2$ quasar sample from the Sloan
Digital Sky Survey. Using simple prescriptions relating quasars to
dark matter halos, we model the observed small-scale clustering
with halo occupation models. Reproducing the large-scale
clustering amplitude requires that the active fraction of the
black holes in the central galaxies of halos is near unity, but
the level of small-scale clustering favors an active fraction of
black holes in satellite galaxies $0.1 \la f_s \la 0.5$ at $z\ga
3$.

%We find tentative evidence that the average duty cycle of
%quasar-hosting satellite halos is $\lesssim 50\%$ at $z\gtrsim 3$.
%On the other hand, the small-scale correlation function favors a
%steeper slope $\gamma>2.5$. If there are no severe selection
%incompleteness in the binary sample, our results then imply either
%a luminosity dependence of quasar clustering at $z\gtrsim 3$, or a
%different halo occupation distribution of quasars from their low
%redshift counterparts.
\end{abstract}

\section{INTRODUCTION}
\label{sec:intro}

%Quasars are among the most luminous objects in the universe, with
%typical bolometric luminosities $\gtrsim10^{45}\,{\rm
%erg\,s^{-1}}$. Although it is generally accepted that the high
%luminosity is induced by gas accretion onto the nuclear
%supermassive black hole (SMBH; Salpeter 1964; Lynden-Bell 1969),
%the mechanisms that trigger such massive nuclear fueling are still
%unclear. Dynamical instabilities, which are required to tunnel a
%significant amount of gas into the central region of the galaxy
%over a relatively short time ($t\ll t_H$) to feed the black hole,
%can result from secular processes (bar/oval instabilities, etc.),
%gravitational tidal interactions during galaxy encounters, or more
%violent direct collisions of galaxies (merger). For the most
%luminous objects ($M_B<-23$, the traditional definition of
%quasars), gas-rich galaxy mergers provide the most efficient way
%of triggering quasar activities (e.g., Hernquist 1989). Indeed,
%this merger scenario is supported by a number of observations,
%such as the ULIRG-quasar connection (e.g., Sanders \& Mirabel
%1996; Canalizo \& Stockton 2001), the evidence of recent mergers
%in quasar hosts (e.g., Bennert \etal\ 2008), and the small-scale
%excess of galaxies around luminous quasars or quasar binaries
%(e.g., Bahcall \etal\ 1997; Serber \etal\ 2006; Hennawi \etal\
%2006; Myers \etal\ 2008). On the theoretical side, a {\em major
%merger}-driven framework is developed by Hopkins \etal\ (2008),
%which nicely explains many of the observational aspects of the
%quasar population and star-burst galaxies.

With the rapid progress in observational and computational
cosmology in the past two decades due to dedicated surveys and
numerical simulations, it is now possible to study the quasar
population within the hierarchical structure formation framework
(e.g., Kauffmann \& Haehnelt 2000; Volonteri, Haardt \& Madau
2003; Wyithe \& Loeb 2003; Hopkins \etal\ 2008; Shankar \etal\
2008, 2009; Shen 2009). If luminous quasars are the progenitors of
the most massive galaxies today, then they occupy the rare peaks
in the initial density fluctuation field, i.e., they are biased
tracers of the underlying matter distribution (e.g., Bardeen
\etal\ 1986; Efstathiou \& Rees 1988; Cole \& Kaiser 1989;
Djorgovski 1999; Djorgovski et al.\ 1999). The quasar two-point
correlation function has now been measured for large survey
samples to unprecedented precision (e.g., Porciani \etal\ 2004;
Croom \etal\ 2005; Myers \etal\ 2006, 2007a; Shen \etal\ 2007,
2008, 2009; da \^{A}ngela \etal\ 2008; Ross \etal\ 2009). These
studies suggest that quasars live in massive dark matter halos of
$M_{\rm halo}\gtrsim\ {\rm a\ few}\times 10^{12}\,h^{-1}M_\odot$;
their bias relative to the underlying matter increases rapidly
with redshift. However, such studies are unable to probe the
smallest scales ($r\lesssim 1\,h^{-1}$Mpc), where matter evolves
nonlinearly and the distributions of quasars within dark matter
halos start to play a role in determining their clustering
properties. This is because fiber-fed multi-object spectroscopic
surveys usually cannot observe two targets closer than the fiber
collision scale $\sim 1'$.

Hennawi \etal\ (2006) compiled a sample of close quasar binaries
at $z<3$ by spectroscopic follow-up observations of candidates
selected from the Sloan Digital Sky Survey (SDSS; York \etal\
2000) imaging data. Using this binary sample, they measured the
correlation function down to scales as small as $R_{\rm prop}\sim
15\,$kpc, where $R_{\rm prop}$ is the transverse separation in
proper units; they confirmed and extended previous tentative
claims (e.g., Djorgovski 1991) that quasars exhibit excess
clustering on small scales (most notably at $R_{\rm prop}\lesssim
40\,{\rm kpc}$) compared with the naive power-law extrapolation of
the large-scale correlation function. This small-scale excess
clustering was confirmed by Myers \etal\ (2007b, 2008) in a more
homogeneous sample, albeit at a lower level of ``excess".

The large-scale quasar correlation function has now been measured
at high redshift ($z\gtrsim3$, Shen \etal\ 2007), where quasars
cluster much more strongly than their low redshift counterparts.
It is natural to extend the work of Hennawi \etal\ (2006) to study
the small-scale quasar clustering at $z>3$. However, such
investigations are challenging for two reasons: first, the number
density of the quasar population drops rapidly after the peak of
quasar activity at $z\sim 2-3$ (e.g., Richards et al.\ 2006);
second, quasar pairs on tens of kpc to $1\,$Mpc scales are rare
occurrences -- only $\lesssim 0.1\%$ of quasars have a close
quasar companion with comparable luminosity. Hence a large search
volume is needed to build up the statistics. Hennawi \etal\ (2009,
hereafter Paper I) have, for the first time, compiled such a
binary quasar sample at $z\gtrsim 3$, which we use here to study
the clustering of quasars on small-scales.

The rareness of close quasar pairs is not in direct contradiction
with the {\em major merger} scenario of quasar triggering, because
the probability that two quasars are triggered and identified
simultaneously during the early stage of a major merger (i.e.,
with separations on halo scales rather than on galactic scales) is
low in theoretical models (e.g., Volonteri et \etal\ 2003; Hopkins
\etal\ 2008). However, even a handful of close quasar pairs will
contribute significantly to the small-scale clustering amplitude
because the mean number density of quasars is so low that the
expected number of random companions on such small scales is tiny.
Also note that although quasar pairs with comparable luminosities
are rare, there might be more fainter companions (i.e., low
luminosity AGN or fainter quasars) around luminous quasars (e.g.,
Djorgovski \etal\ 2007), as expected from the hierarchical merger
scenario.

In this paper we measure the small-scale quasar clustering at
$z\gtrsim 3$ using a set of 15 quasar pairs in the sample of Paper
I. We adopt the same cosmology as in Paper I, with
$\Omega_m=0.26$, $\Omega_\Lambda=0.74$ and $h=0.7$. Comoving units
will be used unless otherwise specified, and we use subscript
$_{\rm prop}$ for proper units.

\section{The Sample}

Our parent sample is the high-redshift binary quasar catalog
presented in Paper I. This sample includes 27 quasar pairs with
relative velocity $|\Delta v|< 2000\,{\rm km\,s^{-1}}$ at
$2.9<z<4.5$, down to a limiting magnitude $i<21$ after correcting
for Galactic extinction, selected over $8142$ ${\rm deg}^2$ of the
SDSS imaging footprint prior to DR6 (Adelman-McCarthy \etal\
2008). The detailed target selection criteria, completeness
analysis, and follow-up spectroscopy can be found in Paper I. To
construct our clustering subsample, we first exclude eight pairs
that failed to pass the selection criteria (for which the
completeness cannot been quantified) described in \S2 of Paper I,
leaving 19 pairs. Second, the follow-up spectroscopic observations
are the most complete out to an angular separation $\theta\approx
60''$, because those targets were assigned higher priority for
follow-up spectroscopy, and therefore we restrict ourselves to
pairs with angular separation $\theta<60''$; this restriction
excludes one pair at $z<3.5$ and three pairs at $z>3.5$. Our final
clustering subsample thus includes 15 pairs with seven pairs at
$z<3.5$ and eight pairs at $z>3.5$, with projected comoving
separations $R\sim 0.1-1\,h^{-1}$Mpc and proper separations
$R_{\rm prop}\sim$ a few tens to a few hundreds of kpc (see fig. 8
of Paper I).

The sparseness of the sample requires different techniques for
measuring the clustering strength, from the traditional binned
$w_p$ statistic (e.g., Davis \& Peebles 1983). Here we adopt the
Maximum-Likelihood (ML) approach used in Shen \etal\ (2009), as
described below. We report our ML estimates and statistical
uncertainties of the small-scale clustering in \S\ref{subsec:ML};
the systematic uncertainties are discussed in
\S\ref{subsec:systematic_error}. To reduce the impact of the
selection incompleteness at $z\sim 3.5$ due to stellar
contaminants (see Paper I), and to explore redshift evolution, we
measure the small-scale clustering in two redshift bins:
$2.9<z<3.5$ (low-$z$) and $3.5<z<4.5$ (high-$z$).

%\begin{figure*}
%  \centering
%    \includegraphics[width=0.48\textwidth]{LH_contour_R0.04-1.00_z2.9-3.5_comp0.6_Agrid.eps}
%    \includegraphics[width=0.48\textwidth]{LH_contour_R0.10-1.25_z3.5-4.5_comp0.7_Agrid.eps}
%    \caption{Likelihood contours for the low-$z$ (left) and high-$z$ (right) cases, where the
%    red contours are the joint $1\sigma$ levels.}
%\end{figure*}

\begin{figure*}
  \centering
    \includegraphics[width=0.48\textwidth]{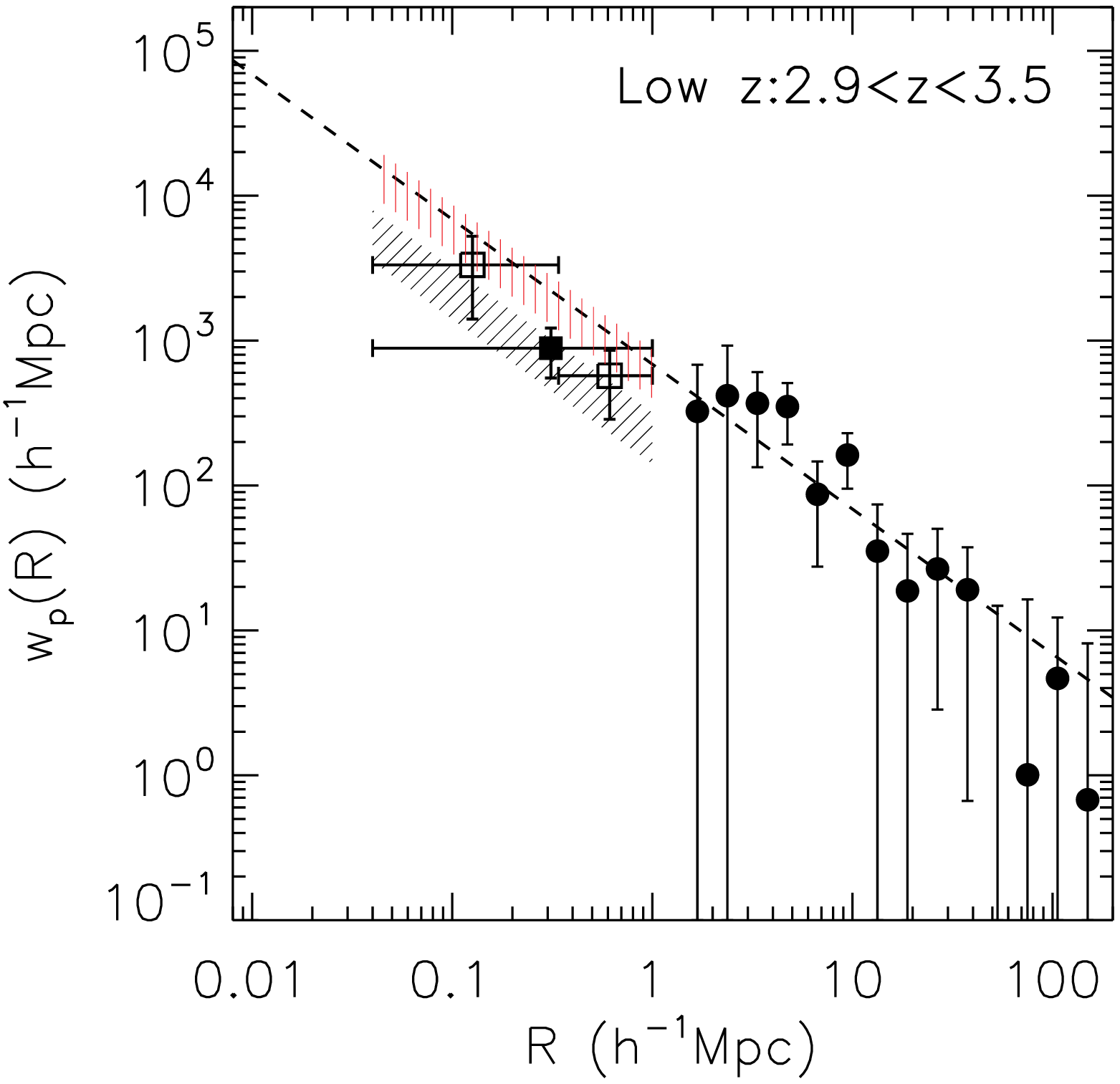}
    \includegraphics[width=0.48\textwidth]{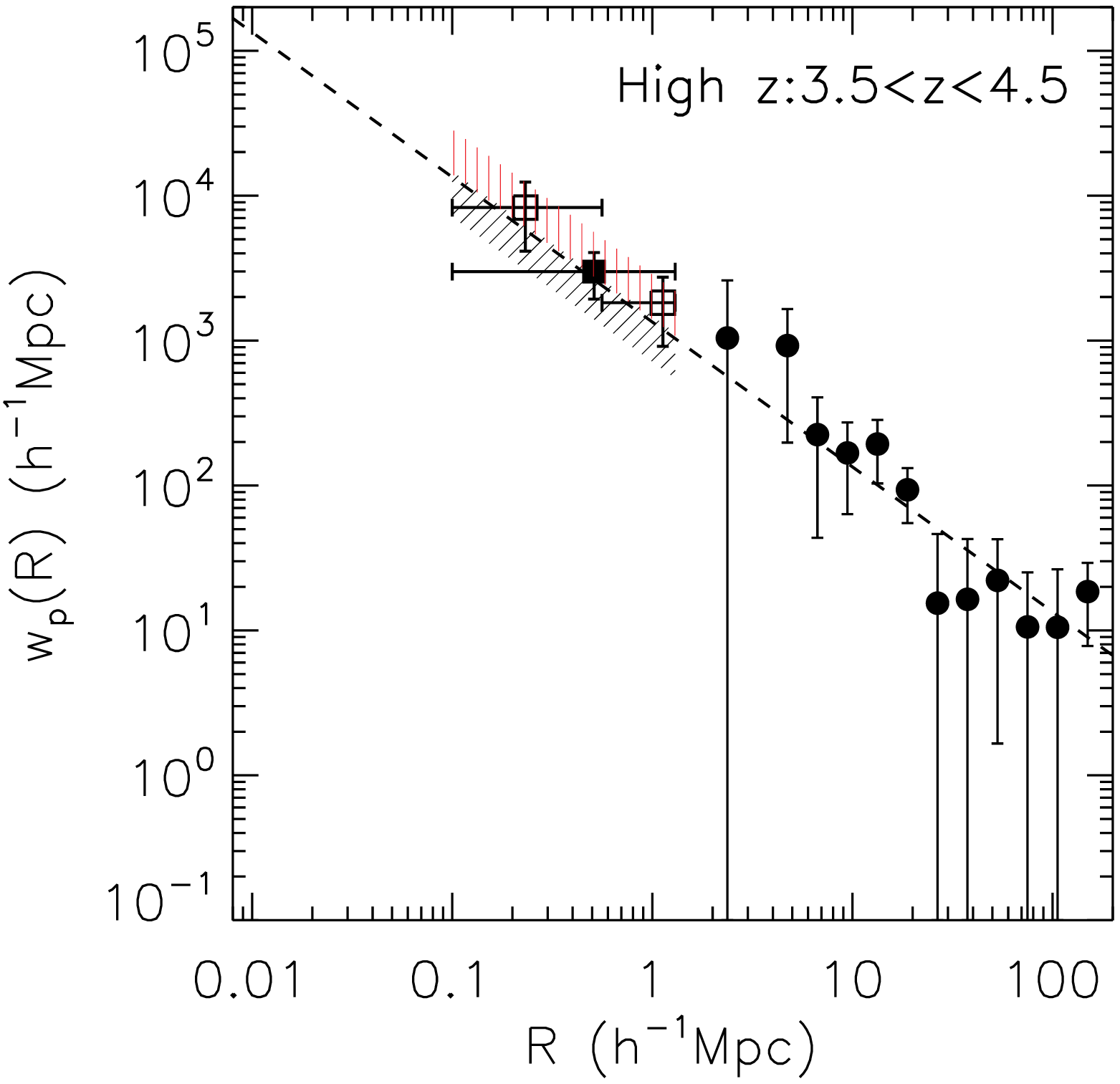}
    \caption{Measurements of the small-scale clustering for the low-$z$
    bin (left) and high-$z$ bin (right). Filled circles are the
    large-scale correlation function data from Shen \etal\
    (2007, the {\em all} sample) and dashed lines are their power-law
    fits with fixed slope $\gamma=2$. Squares are our estimate of $w_p$ using Eqn. (\ref{eqn:wp}),
    estimated in a large radial bin (filled) and two smaller radial bins (open).
    Points are placed at the logarithmic mean of pair separations
    in the bin, horizontal error bars show the bin size,
    and vertical error bars show Poisson errors. The black hatched regions show our ML power-law
    fits to the small-scale pairs (\S\ref{subsec:ML}; $f_{\rm spec}=1$), with the vertical extent enclosing the 1$\sigma$
    statistical uncertainty from the
    ML fitting. If we assume minimal spectroscopic completeness, $f_{\rm spec}=0.38$
    (0.52) for the low-$z$ (high-$z$) bin, the ML results are shown as red hatched regions (see \S\ref{subsec:systematic_error}).
     These estimates, however, should be considered as solid upper limits.} %We overplot three HOD models with different satellite halo duty cycles in
%    blue, cyan and magenta lines for the total (solid) and 1-halo term correlation
%    functions (dotted). See the text in \S\ref{sec:diss} for model details.
    \label{fig:wp}
\end{figure*}

\subsection{Clustering Measurements}\label{subsec:ML}

Here we recast the ML approach of Shen \etal\ (2009). We choose a
power-law model for the underlying correlation function:
$\xi(r)\equiv (r/r_{\rm 0,ML})^{-\gamma_{\rm ML}}$. We then
compute the expected number of quasar pairs within a comoving
cylindrical volume with projected radius $R$ to $R+dR$ and
half-height $\Delta H=20\ h^{-1}$Mpc. This half-height is chosen
to reflect our velocity constraint in defining a quasar pair and
to minimize the effects of redshift distortions and errors.
Assuming Poisson statistics, the likelihood function can be
written as:
\begin{equation}
{\cal L}=\prod_i^N e^{-\mu_i}\mu_i\prod_{j\neq i}e^{-\mu_j}\ ,
\end{equation}
where $\mu=2\pi Rh(R)\,dR$ is the expected number of pairs in the
interval $dR$, the index $i$ runs over all pairs in the sample and
the index $j$ runs over all the elements $dR$ in which there are
no pairs. The expected pair surface density $h(R)$ is given by
\begin{equation}\label{eqn:hr}
h(R)=\displaystyle \frac{1}{2}\int_{z_{\rm min}}^{z_{\rm max}}
f_{\rm comp}(z)n^2(z)\,dV_c\int_{-\Delta H}^{\Delta
H}[1+\xi(\sqrt{R^2+H^2})]dH\ ,
\end{equation}
where $n(z)$ is the cumulative quasar luminosity function down to
a limiting magnitude (in this case $i=21$), $f_{\rm comp}(z)$ is
the completeness in selecting binary candidates for follow-up
spectroscopy as quantified in Paper I (see their fig. 7), and
$V_c$ is the comoving volume between redshifts $z_{\rm min}$ and
$z_{\rm max}$ covered by the binary survey.
%(assumed to be
%constant for either of the two redshift bins, i.e., $f_{\rm
%comp}=0.36$ and $0.5$ for the lower and higher redshift bins
%respectively, Hennawi \etal\ 2009).
The factor of $1/2$ in eqn. (\ref{eqn:hr}) removes duplicate
counts of pairs.

Note we only consider the completeness in target selection $f_{\rm
comp}\equiv f_{\rm targ}$ (i.e., the fraction of quasar binaries
that would have been selected by the algorithm in Paper I)
throughout this section. We discuss the effects of the
completeness of the spectroscopic follow-up of survey candidates
(i.e., the fraction of targets that have been observed), $f_{\rm
spec}$, in \S\ref{subsec:systematic_error}.

Defining the usual quantity $S\equiv -2\ln{\cal L}$ we have
\begin{equation}
S \equiv -2\ln {\cal L} = 2\int_{R_{\rm min}}^{R_{\rm max}} 2\pi
Rh(R)\, dR - 2\sum_i^N\ln[h(R_i)]\ ,
\end{equation}
with all the model-independent additive terms removed. Here
$[R_{\rm min} ,R_{\rm max}]$ is the range of comoving scales over
which we search for quasar pairs. To include all observed pairs
with angular separation
%\footnote{We have tested with including
%pairs with $\theta>60''$ in the ML fitting, and found consistent
%results.}
$\theta<60''$, we choose $[R_{\rm min}, R_{\rm max}]=[0.04,1]\
h^{-1}$Mpc for the low-$z$ bin and $[R_{\rm min}, R_{\rm
max}]=[0.1,1.3]\ h^{-1}$Mpc for the high-$z$ bin. We verified that
our results were not sensitive to the exact values of these
limits. If we fit both $r_{\rm 0, ML}$ and $\gamma_{\rm ML}$ we
found that the best-fit model favors $\gamma_{\rm ML}>2.3$ for
both redshift bins, a value substantially steeper than the slope
on large scales, $\gamma\sim 2$ (e.g., Shen \etal\ 2007). However,
the spectroscopic completeness, $f_{\rm spec}$, probably depends
on angular separation, because we tended to observe the closest
candidates first; this may introduce an artificially steep slope
in the correlation function. Therefore we fix the slope
$\gamma_{\rm ML}=2$ (i.e., close to the measured slope of the
large-scale correlation function, Shen \etal\ 2007) and minimize
the merit function $S$ with respect to $r_{\rm 0,ML}$ only. A
power-law slope $\gamma\sim 2$ is also found for the clustering of
SDSS LRGs to $z\sim 0.4$ (e.g., Masjedi \etal\ 2006) and
photometric SDSS quasars (e.g., Myers et al.\ 2006, 2007a) over a
wide range of scales down to $r=0.01\,h^{-1}$Mpc .

Alternatively, we may estimate the projected correlation function,
i.e., the $w_p$ statistic, for these pairs. Following the
definition of $w_p$ (e.g., Davis \& Peebles 1983), we have
\begin{eqnarray}\label{eqn:wp}
w_p(R)&=&\displaystyle 2\int_0^{\infty} dH\xi_{2D}(R,H)=\displaystyle 2\int_0^{\infty}\bigg(\frac{DD}{RR}-1\bigg)dH \nonumber\\
%&\approx 2\sum\frac{DD}{RR}dH
&\approx& \displaystyle 2dH\frac{\sum DD}{RR}
%&=\frac{2\Delta HN_{\rm pair}}{\frac{1}{2}(f_{\rm comp}\int
%n^2dV_c) 2\pi
%R\Delta R\Delta H} \nonumber \\
%&\approx\frac{2N_{\rm pair}}{(f_{\rm comp}\int n^2dV_c)\pi R\Delta
%R} \nonumber \\
\approx\displaystyle\frac{4N_{\rm pair}}{(\int_{z_{\rm
min}}^{z_{\rm max}} f_{\rm comp}n^2dV_c)\pi(R_2^2-R_1^2)}\ ,
\end{eqnarray}
where $\pi(R_2^2-R_1^2)$ is the projected comoving area of the
cylindrical annulus over which we search for pairs, $R$ is the
geometric mean pair separation in the bin, $n(z)$ is the
cumulative quasar number density, $N_{\rm pair}\equiv \Sigma DD$
is the observed number of quasar pairs in the bin, and $RR$ is the
expected number of random-random pairs in the cylindrical shell
with radii $(R_1,R_2)$ and height $dH$. Note that there are some
approximations and ambiguities involved in Eqn. (\ref{eqn:wp}),
such as the position of the bin center, hence it can only be
treated as a crude estimate for $w_p$.

For both the ML approach and the $w_p$ statistic we need to
estimate the integral $\int_{z_{\rm min}}^{z_{\rm max}} f_{\rm
comp}n^2dV_c$. This requires knowledge of the faint end of the
luminosity function ($i<21$) of quasars at redshift $2.9<z<4.5$.
We have searched the literature for usable LF within these
redshift and luminosity ranges (e.g., Wolf \etal\ 2003; Jiang
\etal\ 2006; Richards \etal\ 2006; Hopkins \etal\ 2007). The Jiang
\etal\ (2006) LF data probe sufficiently faint but do not extend
to $z>3.6$; the Richards \etal\ (2006) data have the desired
redshift coverage but do not probe deep enough. By comparing the
Richards \etal\ LF with the COMBO-17 LF (Wolf \etal\ 2003), we
found that the COMBO-17 PDE fit gives better estimates of the LF
at $z>3.5$, e.g., it agrees well with the Richards \etal\ LF at
the high luminosity end, and produces the expected flattening at
fainter luminosities. Motivated by these comparisons, we adopt a
combination of the Jiang \etal\ fit (at $z<3.5$) and the COMBO-17
PDE fit (at $z>3.5$) for the model LF, scaled to our standard
cosmology. We estimate an uncertainty in the cumulative number
density ($i<21$) of $\sim 20\%$, based on the statistical
uncertainties in these LF fits and comparison between these
optical LFs and the bolometric LF compiled by Hopkins \etal\
(2007), where the faint end LF at these redshifts is further
constrained by X-ray data. This estimate of uncertainty in the
model LF is, however, conservative at $z>3.5$, since there are no
direct optical LF measurements down to $i=21$ within this redshift
range. We will discuss the contribution of the uncertainty in the
LF to the systematic errors in our small-scale clustering
measurements, in \S\ref{subsec:systematic_error}.

%\begin{figure*}
%  \centering
%    \includegraphics[width=0.45\textwidth]{LFcomp_fig20.eps}
%    \includegraphics[width=0.45\textwidth]{cumLF_i21_ratio.eps}
%    \caption{Luminosity function comparisons. {\em Left:} A reproduction of fig. 20 in
%    Richards \etal\ (2006) with the Jiang \etal\ and the COMBO-17 results over-plotted. The
%    Richards \etal\ fit (variable power-law model) underestimates the number density at
%    $3< z< 4.5$ (see their fig. 20); while the extrapolation of the Jiang \etal\
%    fits above $z > 3.6$ clearly over-predicts the number density. For $z<3.5$, the Jiang \etal\
%    LF is the most reliable one to use, while at $z>3.5$ the COMBO-17 PDE LF is our best guess. {\em Right:}
%    the ratios of the cumulative LF at $i<21$ using extrapolations of various fits to our fiducial
%    LF ($z<3.5$: the Jiang \etal\ LF; $z>3.5$: the COMBO-17 PDE LF). Based on the figure, our best guess
%    of the uncertainties in the cumulative number density is $\sim 20\%$.}
%    \label{fig:LFs}
%\end{figure*}

Our clustering measurements are summarized in Fig. \ref{fig:wp},
where we plot for comparison the large-scale ($R\gtrsim 2\
h^{-1}$Mpc) correlation function data from Shen \etal\ (2007, the
{\em all} sample), for the low-$z$ (left) and high-$z$ (right)
bins respectively. The ML approach yields $r_{\rm
0,ML}=8.31^{+1.77}_{-1.61}\ h^{-1}$Mpc for the low-$z$ bin,
%$r_{\rm 0,ML}=6.86^{+1.62}_{-1.47}\ h^{-1}$Mpc and $r_{\rm 0,ML}=14.65^{+3.27}_{-2.90}\ h^{-1}$Mpc
and $r_{\rm 0,ML}=18.22^{+3.47}_{-3.12}\ h^{-1}$Mpc for the
high-$z$ bin, where errors are $1\sigma$ statistical only; these
results are shown as black hatched regions whose horizontal and
vertical extent encloses the fitting range and statistical errors.
For the binned $w_p$ statistic, we take all the pairs and use Eqn.
(\ref{eqn:wp}) to estimate $w_p$ for the two redshift bins with
Poisson errors. We then plot the $w_p$ estimates at the
(geometric) mean values of separations $\langle R\rangle$ as
filled squares in Fig. \ref{fig:wp}. To indicate the uncertainties
in the bin center, we draw horizontal error bars which enclose the
fitting ranges in the ML approach. In both redshift bins we
further divide the pairs into two radial bins (with more or less
equal number of pairs each), with the dividing scale $R=0.34$ and
$0.56\,h^{-1}$Mpc for the low-$z$ and high-$z$ cases (the dividing
scale is set by the geometric mean of the maximum separation of
{\em observed} pairs in the inner bin and the minimum separation
of {\em observed} pairs in the outer bin). The $w_p$ estimates for
the divided $R$ bins are shown in open squares in Fig.\
\ref{fig:wp}. The results of the $w_p$ statistic are consistent
with the ML results within the errors. However, due to the
ambiguity of placing bin centers when there are only a few pairs,
the $w_p$ data points cannot be used in the power-law fit. Our ML
approach is not subject to such ambiguities, and therefore
provides reliable clustering measurements. We tabulate the ML
results in Table \ref{table:r0}.

\begin{deluxetable}{lccc}
\tablecolumns{4} \tablewidth{0pc} \tablecaption{Estimates of $r_0$
for fixed power-law ($\gamma=2$) correlation functions
\label{table:r0}} \tablehead{ & $r_{\rm 0,ML}$ ($f_{\rm spec}=1$)
& $r_{\rm 0,ML}$ (lowest $f_{\rm spec}$) & $r_0$ (large-scale)\\
& $h^{-1}$Mpc & $h^{-1}$Mpc & $h^{-1}$Mpc} \startdata
low-$z$\dotfill  & $8.31^{+1.77}_{-1.61}$  & $13.81^{+2.82}_{-2.52}$ & $14.79\pm 2.12$ \\
\\
high-$z$\dotfill & $18.22^{+3.47}_{-3.12}$ & $25.43^{+4.76}_{-4.28}$ & $20.68\pm 2.52$ \\
\enddata
\tablecomments{The second column lists our ML results assuming
spectroscopic completeness $f_{\rm spec}=1$ (\S\ref{subsec:ML}).
The third column lists ML upper limits assuming the lowest $f_{\rm
spec}$ (\S\ref{subsec:systematic_error}). The fourth column lists
the large-scale correlation lengths from Shen et al.\ (2007, the
{\em all} sample). Uncertainties are 1$\sigma$ statistical only. }
\end{deluxetable}

%At face value, the ML results have comparable or lower clustering
%amplitude at $0.01\lesssim R\lesssim 1\ h^{-1}$Mpc compared with
%the extrapolations from the fits for the large scale correlation
%functions (Shen \etal\ 2007, 2009). However, there are several
%complications with this comparison: 1) it is not justified to use
%the same power-law slope to extrapolate to small scales; 2) the
%quasar sample in Shen \etal\ (2007) has $i<20.2$, while our binary
%sample has $i<21$, thus luminosity-dependent clustering at such
%high redshift and luminosity ranges might play a role (e.g., Shen
%2009); 3) the model luminosity function we are using has
%significant uncertainties, and our target selection completeness
%has a relative uncertainty $\lesssim 10\%$ (Paper I); and finally 4)
%our spectroscopic completeness $f_{\rm spec}$ is not included in
%the modeling, hence all correlation function measurements
%presented here should be considered as lower limits. We discuss
%the systematic uncertainty of our measurements in the next
%section, and the implications for quasar models in
%\S\ref{sec:diss}.

\begin{figure*}
  \centering
    \includegraphics[width=0.48\textwidth]{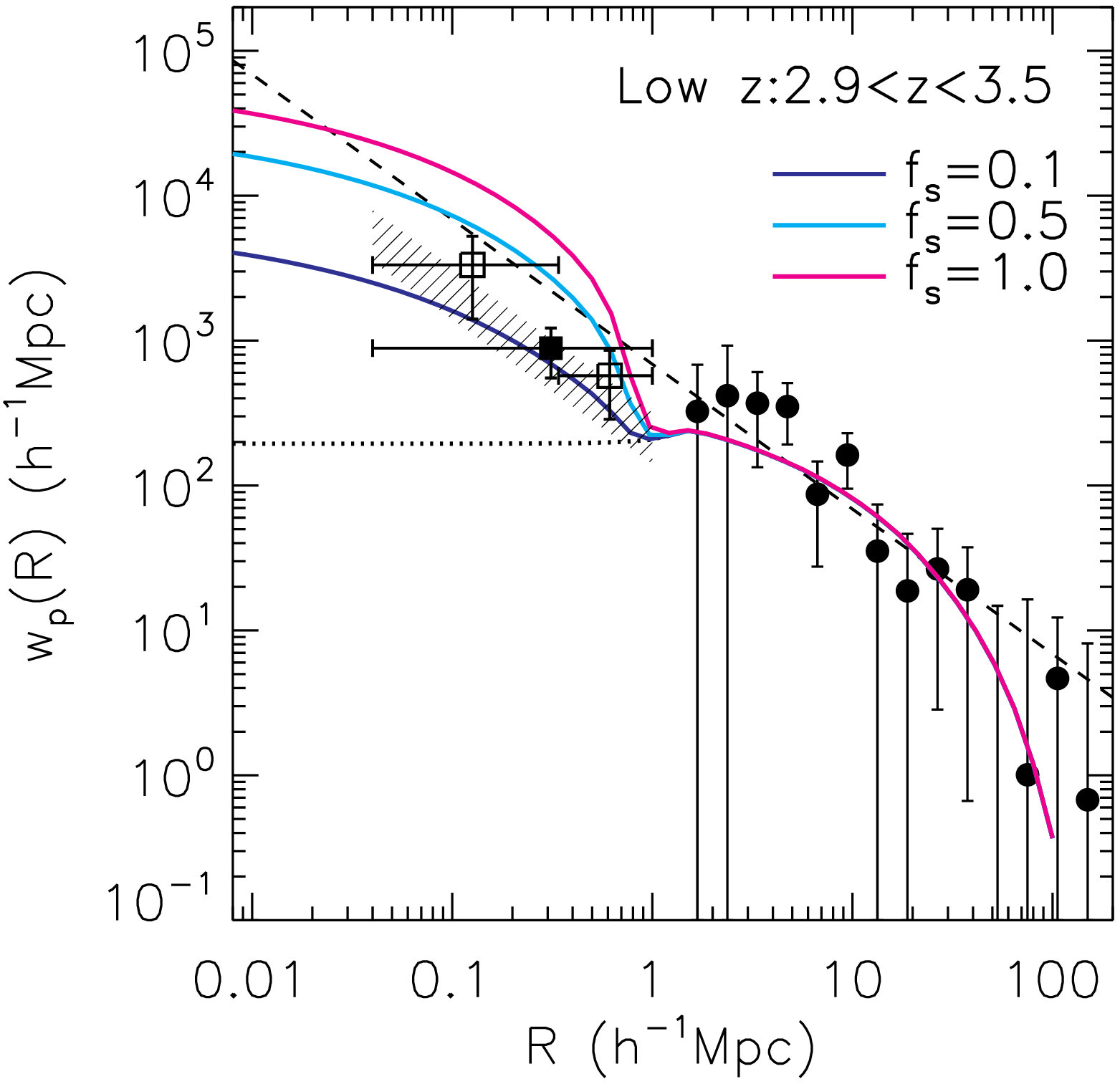}
    \includegraphics[width=0.48\textwidth]{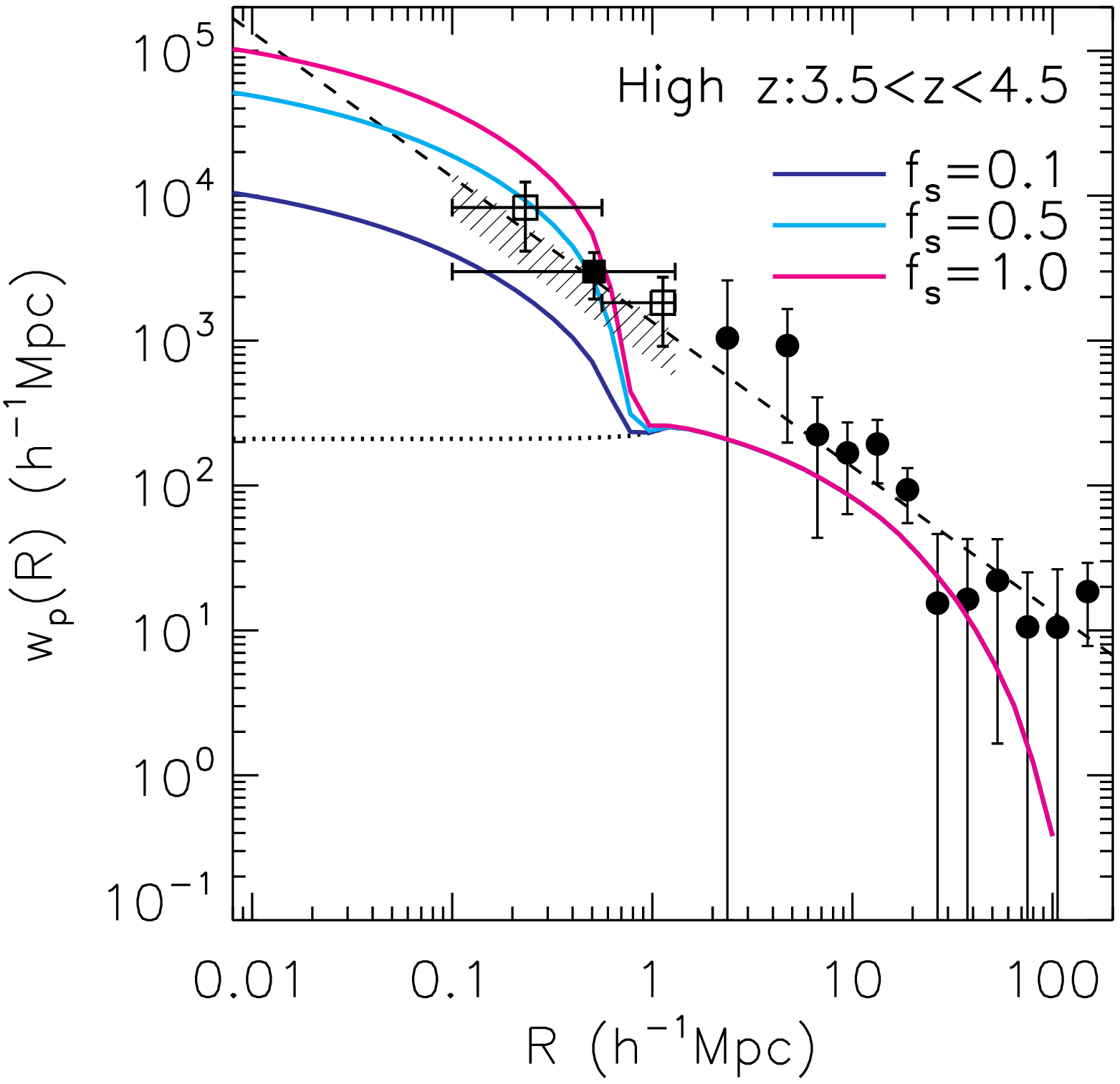}
    \caption{HOD model predictions for a flux limit of $i<21$ compared with the clustering
    data (notations are the same as Fig.\ \ref{fig:wp}).
    The solid lines are the HOD predictions, where the dotted lines are the
    two-halo term contribution to $w_p$. Three HOD models with satellite halo duty cycle
     $f_s=0.1$ (blue), $0.5$ (cyan) and $1.0$ (magenta) are presented. For clarity, we have removed
     the upper limits on the small-scale clustering shown in Fig.\ \ref{fig:wp} (the red hatched regions).}
    \label{fig:wp_HOD}
\end{figure*}

\subsection{Systematic Uncertainties}\label{subsec:systematic_error}

Here we give some quantitative estimation of the systematic
uncertainties in our ML results. The two major systematics come
from the adopted luminosity function and the sample completeness.
Our model luminosity function is quite uncertain down to $i=21$,
especially at $z>3.5$ where no direct optical LF data are
available. As we described above, the uncertainty in the LF is
$\sim 20\%$. In addition, the relative uncertainty in our pair
target selection completeness is $\lesssim 10\%$ (Paper I). These
taken together, introduce a systematic uncertainty in the best-fit
$r_{\rm 0, ML}$ of $\sigma_{r_{\rm 0}}=\pm 1.5\,h^{-1}$Mpc and
$\pm 3.1\,h^{-1}$Mpc for the
%$\sigma_{r_{\rm 0}}=[-1.27,1.88]\,h^{-1}$Mpc and $[-2.54,3.79]\,h^{-1}$Mpc
low-$z$ and high-$z$ bins, respectively; these values are
comparable to the statistical uncertainties reported above.

%This will introduce an additional uncertainty in the best-fit
%$r_{\rm 0,ML}$: $\sigma_{1,r_{\rm 0}}=[-1.26,+1.85]$ and
%$[-2.23,+3.31]$, for the two redshift bins.
%
%The relative uncertainties in our target selection completeness is
%$\sim 10\%$, i.e., $f_{\rm comp}=0.36\pm 0.036$ and $0.5\pm0.05$
%for the low-$z$ and high-$z$ bins respectively. This introduces an
%additional uncertainty in the best-fit $r_{\rm 0,ML}$:
%$\sigma_{2,r_{\rm 0}}=[-0.36,+0.39]$ and $[-0.72,+0.59]$ , for the
%two redshift bins.

In addition, our spectroscopy is incomplete even at $\theta<60''$
-- only $\sim 38\%$ and $\sim 52\%$ of the high-priority low-$z$
and high-$z$ binary targets have been observed (see Table 3 of
Paper I). Therefore we are undoubtedly missing some quasar pairs
and our ML results are lower limits. Because targets further away
from the stellar locus were assigned higher priority (Paper I), it
is difficult to assess the {\em effective} spectroscopic
completeness (because the most promising candidates were observed
first); we expect that the effective spectroscopic completeness is
larger than $50\%$. In the extreme case $f_{\rm spec}=0.38$
(low-$z$) and $0.52$ (high-$z$), we repeat our ML analysis in
\S\ref{subsec:ML} with $f_{\rm comp}=f_{\rm targ}\times f_{\rm
spec}$ and find $r_{\rm 0,ML}=13.81^{+2.82}_{-2.52}\,h^{-1}$Mpc
and $r_{\rm 0,ML}=25.43^{+4.76}_{-4.28}\,h^{-1}$Mpc for the
low-$z$ and high-$z$ case respectively, where errors are 1$\sigma$
statistical. These estimates are shown as red hatched regions in
Fig.\ \ref{fig:wp} and should be considered as solid upper limits.

The ML results in \S\ref{subsec:ML} have comparable or lower
clustering amplitude at $0.1\lesssim R\lesssim 1\ h^{-1}$Mpc than
the extrapolations from the fits for the large-scale correlation
functions (Shen \etal\ 2007, 2009). This does not directly
contradict the results in Hennawi \etal\ (2006) for $z<3$ quasars
since: 1) our sample barely probes scales below $R\sim
0.1\,h^{-1}$Mpc where most of the excess clustering occurs for the
$z<3$ sample (Hennawi \etal\ 2006), and 2) the quasar sample in
Shen \etal\ (2007) has $i<20.2$, while our binary sample has
$i<21$, thus luminosity-dependent clustering at such high redshift
and luminosity ranges might play a role (e.g., Shen 2009). In the
next section we show how these small-scale clustering measurements
can be used to constrain halo occupation models.

%; 3) the model luminosity function we are using has
%significant uncertainties, and our target selection completeness
%has a relative uncertainty $\lesssim 10\%$ (Paper I); and finally 4)
%our spectroscopic completeness $f_{\rm spec}$ is not included in
%the modeling, hence all correlation function measurements
%presented here should be considered as lower limits. We discuss
%the systematic uncertainty of our measurements in the next
%section, and the implications for quasar models in
%\S\ref{sec:diss}.

\section{Discussion}
\label{sec:diss}

%\textbf{Issues to be addressed in the Discussion section. 1.
%luminosity-dependent clustering. 2. non-linear clustering of
%massive halos at high redshift.}

The small-scale clustering measurements presented above can be
used to constrain the statistical occupation of quasars within
dark matter halos at $z\gtrsim 3$. Given that we have a poor
understanding of the physics of quasar formation, we use a simple
phenomenological model relating quasars to halos to model the
observed clustering results. The details of the model will be
presented elsewhere (Shankar \etal, in preparation); below we
briefly describe the model assumptions.

We assume there is a monotonic relationship between quasar
luminosity and the mass of the host dark matter halo (including
subhalos), with a log-normal scatter $\Sigma$ (in dex). Therefore
for a flux-limited quasar sample, the minimal halo mass $M_{\rm
min}$ and the average duty cycle $f$, defined as the fraction of
halos that host a quasar above the luminosity threshold at a given
time, can be jointly constrained from abundance matching and the
clustering strength (e.g., Martini \& Weinberg 2001; Haiman \& Hui
2001; Shen \etal\ 2007; White \etal\ 2008):
\begin{eqnarray}\label{eqn:abund}
n_{{\rm QSO},i<21}(z)&=&\int_{M_{\rm min}}^{\infty}
f(M,z)\Phi_{\rm halo}(M,z)
\nonumber\\
&\times & {\rm erfc}\bigg[\ln\bigg(\frac{M_{\rm
min}}{M}\bigg)\frac{1}{\sqrt{2}\ln(10)\Sigma}\bigg]d\log M\ ,
\end{eqnarray}
where $n_{{\rm QSO},i<21}(z)$ is the cumulative quasar number
density with flux limit $i<21$, $M$ is the halo mass, $\Phi_{\rm
halo}(M,z)$ is the halo mass function per $\log M$ interval, and
$0<f(M,z)<1$ is the average halo duty cycle, which may be a
function of both redshift and halo mass.

In general the halo mass function $\Phi_{\rm halo}(M,z)$ includes
contributions from both halos ($\Phi_c$) and their subhalos
($\Phi_s$), where we use the Sheth \& Tormen (1999) halo mass
function for the former and the {\em unevolved} subhalo mass
function from Giocoli \etal\ (2008) for the latter. It is
important to use the unevolved mass (i.e., mass defined at
accretion before tidal stripping takes place) for subhalos, since
subhalos will lose a substantial fraction of mass during the
orbital evolution within the parent halo. We denote the average
duty cycles for central and satellite halos as $f_c$ and $f_s$
respectively. Note that we assume halos and subhalos of the same
mass host quasars of the same luminosity -- of course, subhalos
within a given halo will be less massive and thus host quasars
fainter on average than the central quasar. The satellite duty
cycle $f_s$ is the fraction of black holes in subhalos that are
active at a given time. The fraction of luminous quasars that are
satellites is always small, regardless of $f_s$, because the
number of massive satellite halos is itself small.

%such that
%\begin{equation}
%f_c\Phi_c + f_s\Phi_s = U(\Phi_c + \Phi_s)\equiv U\Phi_{\rm halo}\
%.
%\end{equation}
%For simplicity we assume constant $U$ and $Q\equiv f_s/f_c$ for
%the mass and redshift ranges considered, and neglect the scatter
%in Eqn. (\ref{eqn:abund}), i.e., $\Sigma=0$. Our model is
%therefore entirely characterized by $U$ and $Q$.

%Realizing the fact that the fraction of binary quasars with
%projected separations $R_{\rm prop}\sim $ tens to hundreds
%$h^{-1}$kpc is extremely low ($\lesssim 0.1\%$ based on the
%observed pairs and cumulative quasar abundance from the model LF),
%quasars predominantly live in central halos. This is expected,
%since the probability of finding massive $M>M_{\rm min}$ satellite
%halos is very low for the relevant mass and redshift ranges.
An important consequence of the rareness of binary quasars is that
the abundance matching, i.e., Eqn. (\ref{eqn:abund}), can be done
using central halos only, and we have $f\approx f_c$, $\Phi_{\rm
halo}\approx \Phi_c$ in Eqn. (\ref{eqn:abund}); the satellite duty
cycle $f_s$ will only affect the small-scale clustering strength.
In order to simultaneously match the large-scale clustering of
$z\gtrsim 3$ quasars (Shen \etal\ 2007, 2009) and their abundance,
Shankar \etal\ (2008, 2009) found large values of duty cycle
$f_c\sim 0.5-1$ are needed, as well as small scatter for the
quasar-halo correspondence, if the Sheth \etal\ (2001) bias
formula is used (cf. Shen \etal\ 2007 for the usage of alternative
bias formulae). For simplicity we fix $f_c=1$ and $\Sigma=0.03$
dex in what follows, which produces adequate fits for the
large-scale clustering and abundance matching\footnote{Although
the model still underpredicts the large-scale clustering a bit for
the high-$z$ bin even with $f_c=1$, as noted in earlier papers
(White \etal\ 2008; Shankar \etal\ 2008; Shen 2009).}.

%Exploring in the $[U,Q]$ parameter space we find
%an overall good fit can be achieved when $U\sim 0.3-0.5$ and
%$Q\sim 0.3-0.5$.
Using Eqn. (\ref{eqn:abund}), we determine the minimal halo mass
to be $M_{\rm min}\sim 10^{13}\,h^{-1}M_\odot$ for both redshift
bins. We then use standard halo occupation distribution (HOD)
models (e.g., Tinker \etal\ 2005) to compute the one-halo term
correlation function with different values of satellite duty cycle
$0<f_s<1$.

%The physical constraints on the combination of $U$ and $Q$ are
%$0\le f_c\le 1$ and $0\le f_s\le 1$.

Fig. \ref{fig:wp_HOD} shows several examples of our HOD model at
$z=3.1$ (left panel) and $z=4$ (right panel) with $f_s=0.1$
(blue), $0.5$ (cyan) and $1.0$ (magenta) for a flux limit of
$i=21$. Solid lines are the total correlation while the dotted
line denotes the two-halo term contribution. As expected, the
value of $f_s$ has no effect on the large-scale clustering; it
only changes the small-scale clustering amplitude. These are not
actual fits to the data because the quality of our measurements
does not allow a reliable HOD fit. Nevertheless, it seems that
some active satellite halos are required, but only $\lesssim 50\%$
of satellite halos can be active at a given time in order not to
overshoot the small-scale clustering. This constraint is less
stringent if we consider instead the upper limits on the
small-scale clustering discussed in
\S\ref{subsec:systematic_error}. One potential concern regarding
our model is that the adopted subhalo mass function has not yet
been tested against simulations for the extreme high-mass end and
redshift ranges considered here; nevertheless our model approach
demonstrates how the small-scale clustering measurements can be
used to constrain quasar occupations within halos. We defer a more
detailed investigation on the uncertainties and caveats of our
halo models to a future paper (Shankar \etal, in preparation).

\section{CONCLUSIONS}

We have measured the small-scale ($0.1\,h^{-1}{\rm Mpc}\lesssim
R\lesssim 1\,h^{-1}{\rm Mpc}$) clustering of quasars at high
redshift ($z\gtrsim 3$), based on a sample of $15$ close binaries
from Paper I. Strong clustering signals are detected, comparable
to or lower than the extrapolations from the large-scale
clustering based on SDSS quasar samples. The small-scale
clustering increases in strength from $z\sim 3$ to $z\sim 4$,
consistent with that of the large-scale clustering (Shen \etal\
2007, 2009).

Using a simple prescription relating quasars to dark matter halos,
we constrain the average duty cycles of satellite halos at
$z\gtrsim 3$ from the small-scale clustering measurements. We
found tentative evidence that only $\sim 10\%-50\%$ of satellite
halos with mass $\ga 10^{13}\, h^{-1}M_\odot$ can host an active
quasar (with $i<21$). With the completion of our ongoing binary
quasar survey, we will have better estimates of the spectroscopic
completeness and therefore will confirm our results.

Future surveys of fainter binary quasars at $z>3$ will increase
the sample size and hence the signal-to-noise ratio of the
small-scale clustering measurements. These measurements, together
with better understandings of the halo/subhalo abundance and
clustering at $z>3$ from simulations, will provide important clues
to the formation of quasars at high redshift.

\acknowledgements We thank Silvia Bonoli, Charlie Conroy, Phil
Hopkins and Linhua Jiang for helpful discussions. This work was
partially supported by NSF grants AST-0707266 (YS and MAS) and
AST-0607634 (DPS). FS acknowledges partial support from NASA grant
NNG05GH77G and from the Alexander von Humboldt Foundation. SGD and
AM acknowledge partial support from NSF grant AST-0407448 and the
Ajax Foundation.

\end{document}